
\input harvmac
\input epsf

\def\half{{1 \over 2}}

\def\laplace{{\kern1pt\vbox{\hrule height 1.2pt\hbox{\vrule width 1.2pt\hskip
  3pt\vbox{\vskip 6pt}\hskip 3pt\vrule width 0.6pt}\hrule height 0.6pt}
  \kern1pt}}
\def\scriptlap{{\kern1pt\vbox{\hrule height 0.8pt\hbox{\vrule width 0.8pt
  \hskip2pt\vbox{\vskip 4pt}\hskip 2pt\vrule width 0.4pt}\hrule height 0.4pt}
  \kern1pt}}

\def\p{\phi}
\def\pa{\phi^a}
\def\s{\sigma}
\def\S{\Sigma}
\def\sm{$\sigma$-model~}
\def\mp{m_\phi}
\def\mph{\widehat{m}_\phi}
\def\ms{m_\sigma}
\def\l{\lambda}
\def\L{\Lambda}
\def\O{{\cal O}}

\newcount \pageit
\footline={\tenrm\hss \ifnum\pageit=0 \hfill \else \number\pageno \fi\hss}
\pageit=0
\pageno=0
\font\titlerm=cmr10 scaled \magstep3
\font\titleit=cmti10 scaled \magstep3
\centerline{\titlerm Black Hole Entropy in the {\titleit O(N)} Model}
\vskip 24pt
\centerline{D.~Kabat, S.~H.~Shenker, and M.~J.~Strassler}
\vskip 8pt
\centerline{\it Department of Physics and Astronomy}
\centerline{\it Rutgers University}
\centerline{\it Piscataway, NJ 08855--0849}
\vskip 8pt
\centerline{\tt kabat, shenker, strasslr@physics.rutgers.edu}
\vskip 0.9 true in
\leftskip = 0.5 true in
\rightskip = 0.5 true in
\noindent
We consider corrections to the entropy of a black hole from an $O(N)$
invariant linear $\s$-model.  We obtain the entropy from a $1/N$
expansion of the partition function on a cone.  The entropy arises
from diagrams which are analogous to those introduced by Susskind and
Uglum to explain black hole entropy in string theory.  The
interpretation of the \sm entropy depends on scale.  At short
distances, it has a state counting interpretation, as the entropy of
entanglement of the $N$ fields $\pa$.  In the infrared, the effective
theory has a single composite field $\s \sim \pa \pa$, and the state
counting interpretation of the entropy is lost.
\leftskip = 0.0 true in
\rightskip = 0.0 true in
\vfill
\vskip -28 pt
\noindent hep-th/9506182 \hfill June 1995
\smallskip
\noindent RU--95--34
\eject
\pageit=1

\def\pr{{\it Phys.~Rev.~}}
\def\prl{{\it Phys.~Rev.~Lett.~}}

\def\np{{\it Nucl.~Phys.~}}
\def\pl{{\it Phys.~Lett.~}}
\def\prep{{\it Phys.~Rep.~}}

\def\cqg{{\it Class.~Quant.~Grav.~}}
\def\mpl{{\it Mod.~Phys.~Lett.~}}
\def\ijmp{{\it Int.~J.~Mod.~Phys.~}}
\def\ap{{\it Ann.~Phys.~}}

\nref\SussUg{L.~Susskind, hep-th/9309145\semi
L.~Susskind and J.~Uglum, \pr {\bf D50}, 2700 (1994), hep-th/9401070.}
\nref\Haw{See S.~W.~Hawking, in {\it General Relativity, an Einstein
Centenary Survey}, S.~W.~Hawking and W.~Israel, eds.~(Cambridge, 1979).}
\nref\cones{M.~Ba\~nados, C.~Teitelboim, and J.~Zanelli, \prl {\bf 72},
957 (1994), gr-qc/9309026\semi
S.~Carlip and C.~Teitelboim, gr-qc/9312002.}
\nref\tH{G.~'t Hooft, \np {\bf B256}, 727 (1985).}
\nref\BKLS{L.~Bombelli, R.~K.~Koul, J.~Lee, R.~D.~Sorkin, \pr {\bf D34},
 373 (1986).}
\nref\Sred{M.~Srednicki, \prl {\bf 71}, 666 (1993).}
\nref\CalWil{C.~Callan and F.~Wilczek, \pl {\bf B333}, 55 (1994),
hep-th/9401072.}
\nref\KabStr{D.~Kabat and M.~J.~Strassler, \pl {\bf B329}, 46 (1994),
hep-th/9401125.}
\nref\Dow{J.~S.~Dowker, \cqg {\bf 11}, L55 (1994), hep-th/9401159.}
\nref\HLW{C.~Holzhey, F.~Larsen, and F.~Wilczek, \np {\bf B424} (1994)
433, hep-th/9403108.}
\nref\FPST{T.~Fiola, J.~Preskill, A.~Strominger, and S.~Trivedi,
\pr {\bf D50}, 3987 (1994), hep-th/9403137.}
\nref\DLM{J.-G.~Demers, R.~Lafrance, and R.~C.~Myers, gr-qc/9503003.}
\nref\Kab{D.~Kabat, hep-th/9503016.}
\lref\Israel{W.~Israel, \pl {\bf A57}, 107 (1976).}
\lref\thermal{See for example G.~L.~Sewell, \ap {\bf 141}, 201 (1982)\semi
S.~Fulling and S.~Ruijsenaars, \prep {\bf 152}, 135 (1987).}
\lref\Unruh{W.~G.~Unruh, \pr {\bf D14}, 870 (1976).}
\lref\LarWil{F.~Larsen and F.~Wilczek, hep-th/9408089.}
\lref\Barb{J.~L.~F.~Barbon, \pr {\bf D50}, 2712 (1994), hep-th/9402004.}
\lref\DeAlOh{S.~P.~de Alwis and N.~Ohta, hep-th/9412027\semi
S.~P.~de Alwis and N.~Ohta, hep-th/9504033.}
\lref\Sol{S.~Solodukhin, hep-th/9504022.}
\lref\particle{J.~de Boer, B.~Peeters, K.~Skenderis, and P.~van
Nieuwenhuizen, hep-th/9504097.}
\lref\AGW{See for example L.~Alvarez-Gaum\'e and E.~Witten, \np {\bf B234}
(1983) 269.  Compare equations (73) and (103).}
\lref\BirrDav{See N.~D.~Birrell and P.~C.~W.~Davies, {\it Quantum Fields
in Curved Space} (Cambridge, 1982) section 6.1.}
\lref\dscale{S.~Nishigaki and T.~Yoneya, \pl {\bf B268} (1991) 35\semi
J.~Zinn-Justin, \pl {\bf B257} (1991) 335, hep-th/9112048\semi
P.~Di Vecchia, M.~Kato, and N.~Ohta, \ijmp {\bf A7} (1992) 1391\semi
P.~Di Vecchia and M.~Moshe, \pl {\bf B300} (1993) 49, hep-th/9211132.}
\lref\renorm{S.~Solodukhin, \pr {\bf D51}, 609 (1995), hep-th/9407001\semi
S.~Solodukhin, \pr {\bf D51}, 618 (1995), hep-th/9408068\semi
D.~Fursaev, \mpl {\bf A10}, 649 (1995), hep-th/9408066\semi
D.~Fursaev and S.~Solodukhin, hep-th/9412020.}

\newsec{Introduction}

Within ordinary gravity, the dynamical origins of black hole entropy
have always been obscure.  But gravitation emerges as a low energy
phenomenon from string theory, and Susskind and Uglum have proposed
that, in terms of underlying string degrees of freedom, a state
counting interpretation of black hole entropy is possible \SussUg.

We will discuss black hole entropy within the following framework.
The temperature of a black hole is fixed by requiring that the
Schwarzschild metric, continued to imaginary time, provide a smooth
solution of the Einstein equations.  This forces the periodicity of
the Euclidean time coordinate to be the inverse Hawking temperature.
Entropy is obtained from a partition function by varying with respect
to temperature; in the case of a black hole, this variation introduces
a conical singularity on the horizon.  The classical action for
gravity evaluated on a space with a conical singularity gives rise to
the classical Hawking--Beckenstein entropy \refs{\Haw, \cones}, while
quantum corrections to the classical entropy result from fluctuations
of the metric or matter fields in the background with the conical
singularity.

Through genus one, the string diagrams Susskind and Uglum claim are
responsible for black hole entropy are shown in Fig.~1.  The genus
zero diagram gives rise to the classical $\O(1/\hbar)$
Hawking--Beckenstein entropy of a black hole.  It can be understood as
counting the states of a half-string with its ends stranded on the
horizon.  Two diagrams arise at genus one.  The first represents a
closed string encircling the horizon, and can be understood as
counting the states of a closed string.  The second represents
represents an interaction between a closed string and an open string
stranded on the horizon.  Such interactions can be thought of as
making corrections to the state counting entropy which is present at
genus zero.
\midinsert
\hfil \epsfbox{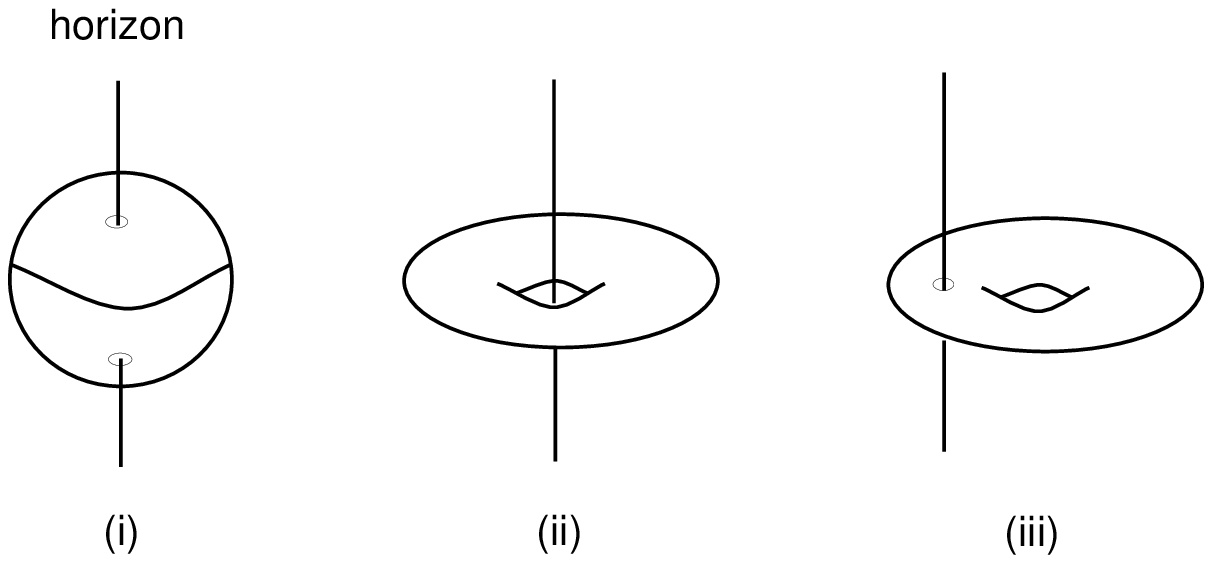} \hfil
\medskip
\centerline{{\bf Fig.~1.}  String diagrams that generate black hole
entropy.}
\bigskip
\hfil \epsfbox{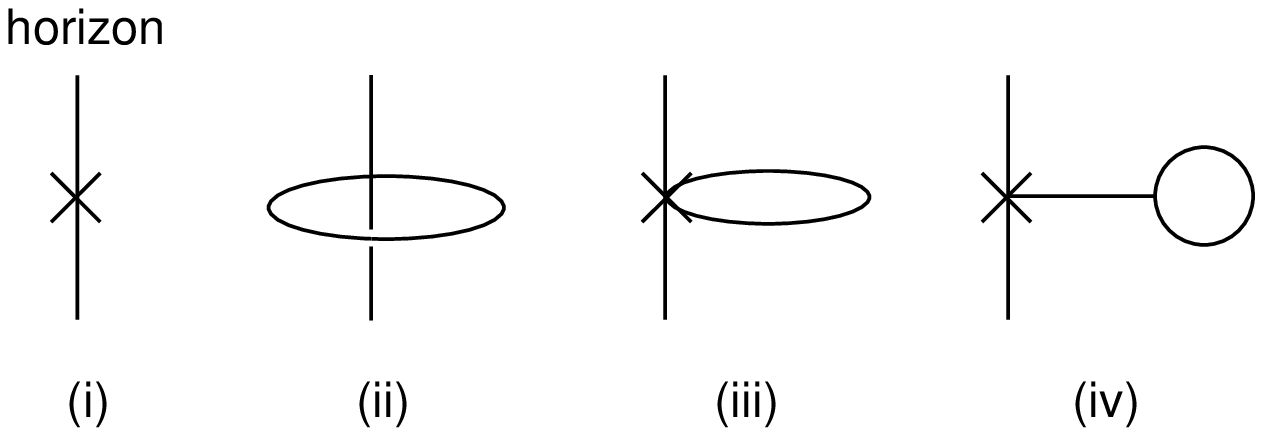} \hfil
\medskip
\centerline{{\bf Fig.~2.}  Low energy limit of the string diagrams.}
\endinsert

This state counting interpretation of black hole entropy is lost at
low energies, when string degrees of freedom are not visible.  At low
energies, the string diagrams reduce to the particle diagrams of
Fig.~2.  Diagram (i) represents a term in the effective action that is
localized on the horizon.  It is responsible for the classical
Hawking--Beckenstein entropy, but has no state counting
interpretation.  Diagram (ii) does have a state counting
interpretation, as counting the states of a particle encircling the
horizon.  In diagrams (iii) and (iv) contact interactions with the
horizon appear.  These contact interactions do not admit state
counting interpretations.  Diagram (iii) can be represented as a path
integral over particle paths that begin and end on the horizon.  In
diagram (iv) a field acquires an expectation value at one loop, which
then couples to the horizon.  In a supersymmetric string theory, all
one point functions on the torus vanish, so this last diagram will not
be present.  In fact one point functions vanish to all orders in the
genus expansion as long as supersymmetry is unbroken, so contributions
of this general form can arise only below the scale of supersymmetry
breaking.

In this paper, we will study corrections to the entropy of a black
hole from an $O(N)$ invariant linear $\s$-model.  For simplicity we
work in two Euclidean dimensions, and consider only infinitely massive
black holes, for which curvature vanishes everywhere on-shell.  The
\sm exhibits some analogous effects within field theory.  In
particular, diagrams appear at short distances in the \sm which are
analogous to the string diagrams in Fig.~1.  In the low energy
effective theory for the $\s$-model, diagrams appear which are
analogous to the particle diagrams in Fig.~2.  Moreover, at short
distances, the \sm corrections to the entropy have a state counting
interpretation as entropy of entanglement.  This state counting
interpretation is lost at low energies.

This paper is organized as follows.  In section 2, we discuss the
relationship between black hole entropy and entropy of entanglement,
and show that this relationship leads to an underlying state counting
interpretation of the \sm contribution to black hole entropy.  We also
discuss the way in which the state counting interpretation of black
hole entropy breaks down for non-minimally coupled fields, such as
appear in the low energy description of the $\s$-model.  In section 3,
we renormalize the \sm to determine its low energy effective action.
In section 4 we compute the entropy of the $\s$-model, and discuss how
its interpretation changes at different length scales.  In section 5
we discuss the implications of our results for string theory.

\newsec{Entropy of Entanglement and Conical Singularities}

In this section we discuss the relationship between corrections to the
entropy of a black hole, obtained from a partition function on a cone,
and entropy of entanglement \refs{\tH {--} \Kab, \SussUg}.  For the
\sm we will argue that these two entropies are identical at short
distances.  This leads to an underlying state counting interpretation
of the \sm contribution to black hole entropy.  We will also show that
their equivalence is broken for fields with a non-minimal coupling to
curvature, such as arise in the low energy description of the
$\s$-model.

We begin by defining entropy of entanglement.  Consider a quantum
field in flat space.  Suppose that it is in its ground state,
described by the pure density matrix $|0><0|$.  Introduce an imaginary
boundary that divides space into two regions, and form the reduced
density matrix $\rho_{\rm red}$ from $|0><0|$ by tracing over all
degrees of freedom located in one of the regions.  Entropy of
entanglement is defined by $S = -\, {\rm Tr}\, \rho_{\rm red} \log
\rho_{\rm red}$.  For simplicity we will only treat the case where
space is divided in half by an imaginary planar boundary.

Entropy of entanglement is ultraviolet divergent in most quantum field
theories, so we must introduce an ultraviolet cutoff $\L$.  The ``full
entropy of entanglement'' of the theory is defined as above in terms
of all the dynamical degrees of freedom present in the theory up to
the cutoff $\L$.  A Wilsonian effective action at a scale $\mu < \L$
is constructed by integrating out the degrees of freedom with momentum
between $\mu$ and $\L$.  We introduce the notion of ``effective
entropy of entanglement,'' that is, the entropy of entanglement
calculated in the effective theory at the scale $\mu$.  When we state
that, for some theory, black hole entropy and entropy of entanglement
are identical, we are referring to the full entropy of entanglement.

Note that the effective entropy of entanglement is less than the full
entropy of entanglement: as degrees of freedom are integrated out, the
effective entropy of entanglement decreases.  In contrast, black hole
entropy is defined in terms of the partition function on a cone, a
quantity which does not change as degrees of freedom are integrated
out.  In the infrared, the difference between the effective and full
entanglement entropies is accounted for by the appearance of terms in
the Wilsonian effective action involving the background curvature.
This is a general phenomenon, which we will explicitly see occur in
the $\s$-model.

We now review the formal argument that relates entropy of entanglement
to black hole entropy.  The argument uses the Rindler Hamiltonian
(generator of Euclidean rotations) $H_R$.  One can show that the
reduced density matrix takes a thermal form in terms of $H_R$, in that
$\rho_{\rm red} = e^{-2\pi H_R}$ \refs{\Israel, \thermal, \SussUg,
\KabStr, \LarWil}.  This is a statement of the Unruh effect \Unruh,
namely, that the Minkowski vacuum state of a field is seen as a
thermal state by a Rindler observer.  The entropy of entanglement of a
field is therefore the same as the thermal entropy which the field
carries in Rindler space, which can be obtained by standard
thermodynamics from the Rindler thermal partition function $Z(\beta) =
{\rm Tr} \, e^{-\beta H_R}$.  But $Z(\beta)$ is the partition function
for the theory on a cone with deficit angle $2\pi - \beta$, so Rindler
thermal entropy is the same as the contribution of the field to the
entropy of a black hole.

This formal argument for equivalence has been tested by explicit
calculation in several theories.  For free scalar and spinor fields,
the full entropy of entanglement, Rindler thermal entropy, and entropy
on a cone are indeed equal \refs{\thermal, \LarWil, \DeAlOh}.  The
equivalence does not always hold, however.  For free vector fields,
singular effects arise at the origin, which make the Rindler
Hamiltonian ill-defined, and cause the equivalence to break down \Kab.

For an interacting minimally coupled scalar field theory the
equivalence holds to all orders in perturbation theory: black hole
entropy and full entropy of entanglement are identical.  In
particular, the equivalence holds for the $\s$-model, and leads to a
state counting interpretation of the \sm contribution to the entropy
of a black hole.  The proof of equivalence is straightforward.  To be
concrete, we consider a scalar field in $1+1$ dimensions with a ${1
\over 4} \lambda \phi^4$ self-coupling.  We work in polar coordinates
$(r, \theta)$ on the Euclidean plane.  The canonical Rindler
Hamiltonian may be constructed in the usual way,
$$
H_R = \int _0^\infty r dr \left( \half \pi^2
 + \half (\partial_r \phi)^2 + \half m^2 \phi^2 + {1 \over 4} \lambda
   \phi^4 \right)
$$
where $\pi$ is the momentum conjugate to $\phi$.

We first show that Rindler thermal entropy is the same as black hole
entropy.  Rindler thermal entropy may be obtained from the partition
function $Z(\beta) = {\rm Tr} \, e^{-\beta H_R}$.  This partition
function can be calculated in perturbation theory, by separating $H_R$
into free and interacting pieces.  The Feynman rules for computing
$Z(\beta)$ are identical to the Feynman rules used to evaluate the
partition function on a cone -- in particular, the two point function
is simply the Green's function on a cone \thermal.  Rindler thermal
entropy and black hole entropy therefore have identical perturbation
series.  The two entropies will be equal provided the two calculations
are cut off in the same way; this can be achieved by using
Pauli-Villars regulator fields as in \DLM.

To complete the proof we show that Rindler thermal entropy is the same
as entropy of entanglement \refs{\Israel, \thermal, \SussUg,
\KabStr, \LarWil}.  Suppose we divide space at the origin $(r=0)$,
and compute the entropy of entanglement of one half of space
($\theta=0$) with the other half ($\theta = \pi$).  This is defined in
terms of the reduced density matrix $\rho_{\rm red}$, constructed from
the vacuum density matrix $|0><0|$ by tracing over all degrees of
freedom located at $\theta = \pi$.  To all orders of perturbation
theory, and for any operator $A$ located at $\theta = 0$, the
expectation values ${\rm Tr} \, (\rho_{\rm red} A)$ and ${\rm Tr} \,
(e^{-2\pi H_R} A)$ are equal, because (as above) the Feynman rules are
identical.  This shows that $\rho_{\rm red} = e^{- 2 \pi H_R}$, which
implies that entropy of entanglement is equal to Rindler thermal
entropy.

We now discuss a theory for which black hole entropy and full entropy
of entanglement are not equivalent.  Consider a scalar field in two
dimensions with a non-minimal coupling to curvature.
$$
S = \int_{\cal M} \!\! d^2x \sqrt{g} \left({1 \over 2} g^{\mu\nu}
      \partial_\mu \phi \partial_\nu \phi + {1 \over 2} m^2 \phi^2
      + {1 \over 2} \xi R \phi^2 \right)
$$
A non-minimal coupling of this form will arise in the low energy
description of the $\s$-model.  To see that the non-minimal coupling
breaks the equivalence, note that entropy of entanglement is defined
in terms of the vacuum state of the field in flat Minkowski space, and
should therefore be insensitive to any terms in the action that vanish
in flat space.  The partition function on a cone, on the other hand,
is affected by a non-minimal coupling to the curvature present at the
tip of the cone \Sol.  This difference breaks the connection between
entropy of entanglement and black hole entropy.

To justify the claim that entropy of entanglement is not affected by
the non-minimal coupling, we work in flat space, and consider the
effect of the non-minimal coupling in greater detail.  In flat space,
the non-minimal coupling only enters via a boundary term, involving
the extrinsic curvature $K$, which we must add to the action to make
the energy-momentum tensor well-defined.
$$
S_{\rm boundary} = \int_{\partial {\cal M}} \!\!\! ds \,\, \xi K \phi^2
$$
The vacuum wavefunctional in infinite flat space can be obtained from
a path integral on a semi-disc of radius $L$, $\{(x,y) : x^2 + y^2 \le
L^2, \, y \ge 0\}$.  As $L \rightarrow \infty$, with the field taken
to vanish on the semi-circular boundary at $x^2+y^2=L^2$, this path
integral produces the desired flat space vacuum wavefunctional on the
boundary at $y=0$.  The bulk curvature vanishes on the semi-disc,
while the extrinsic curvature is $1/L$ along the semi-circular
boundary, with $\delta$-functions of strength $\pi/2$ at the two
corners.  This leads to the vacuum wavefunctional
$$
\psi[\phi] =  {\cal N} \lim_{L \rightarrow \infty} e^{ - \pi \xi
             \left(\phi^2(L,0) + \phi^2(-L,0)\right)/2} \, \psi_0[\phi]
$$
where $\psi_0[\phi]$ is the vacuum wavefunctional for a minimally
coupled field, and ${\cal N}$ is a normalization constant.  We see
that in flat space the non-minimal coupling only affects the vacuum
state at spatial infinity.  Because we consider a massive field,
correlation functions calculated in this vacuum are likewise only
affected at spatial infinity.

What is the entropy of entanglement of this state?  Suppose we divide
space at $x=0$.  For a massive minimally coupled field the degrees of
freedom at $x = \pm L$ are exponentially uncorrelated with the degrees
of freedom located across the division.  From the form of the
wavefunctional we see that they remain uncorrelated in the presence of
the non-minimal coupling; it follows that the non-minimal coupling
does not affect the entropy of entanglement.

The fact that the correlation functions are only affected at spatial
infinity has an important consequence.  It implies that the Rindler
Hamiltonian which generates angular evolution in flat space is not
affected by the non-minimal coupling, aside from a possible surface
term at spatial infinity.  Even with a non-minimal coupling, one can
regard entropy of entanglement as counting the degrees of freedom of a
well-defined dynamical system governed by this flat space Rindler
Hamiltonian.

Note that, for non-zero $\xi$, the partition function on a cone is
{\it not} generated by the flat space Rindler Hamiltonian -- in fact
it seems unlikely that the partition function on a cone can be
generated by a Hermitian angular Hamiltonian at all, for reasons given
below.  We can now state more precisely why the equivalence between
entropy of entanglement and black hole entropy breaks down for a
non-minimally coupled field.  The Rindler Hamiltonian is perfectly
well-defined, and the reduced density matrix is given by $\rho_{\rm
red} = e^{-2 \pi H_R}$, so entropy of entanglement is equal to Rindler
thermal entropy.  But it is not equal to black hole entropy, because
$H_R$ does not generate the partition function on a cone.  This
provides an interesting contrast to the way in which the equivalence
is broken for a vector field \Kab.  For a vector field, singular
effects present at the origin make the Rindler Hamiltonian
ill-defined, even in flat space.

We now study the behavior of the same non-minimally coupled field in
curved space.  To compute the partition function on a cone, and to
gain more insight into the effect of the non-minimal coupling, we
introduce a particle path integral representation for the partition
function.  We work in two Euclidean dimensions, with a proper time
cutoff $\epsilon$ as a regulator.
\eqn\part{\eqalign{
\beta F &= \half \log \det \left(-\laplace + m^2 + \xi R\right)\cr
        &= - \half \int_{\epsilon^2}^\infty {ds \over s}
             \int\limits_{\,\,\,\,x(\tau+s)=x(\tau)}\!\!\!\!\!\!
             {\cal D}x(\tau)\,\, \exp \left[-\int_0^s d\tau\,
             \left({1 \over 4} g_{\mu\nu}\dot{x}^\mu \dot{x}^\nu
               + m^2 + \xi R(x)\right)\right]\cr}}
A careful definition of the particle path integral would have
additional terms present in the action, beyond those we have indicated
here \particle.  The naive action we use here is sufficient for our
purposes, however, as we only wish to show that the effect of the
non-minimal coupling is to produce an additional contact term in the
partition function.  To see this, we expand the free energy in powers
of the non-minimal coupling, following \Sol.  The scalar curvature on
a cone is $R = 2 (2\pi-\beta) \delta^2(x) / \sqrt{g}$.
$$\eqalign{
\beta F &= - \half \int_{\epsilon^2}^\infty {ds \over s}
             \int\limits_{\,\,\,\,x(\tau+s)=x(\tau)}\!\!\!\!\!\!
             {\cal D}x(\tau)\,\, \exp \left[-\int_0^s d\tau\,
             \left({1 \over 4} g_{\mu\nu}\dot{x}^\mu \dot{x}^\nu
               + m^2\right)\right]\cr
   &\qquad + \xi \, (2\pi-\beta)\int_{\epsilon^2}^\infty ds
             \int\limits_{\,\,\,\,x(0)=x(s)=0}\!\!\!\!\!\!
             {\cal D}x(\tau)\,\, \exp \left[-\int_0^s d\tau\,
             \left({1 \over 4} g_{\mu\nu}\dot{x}^\mu \dot{x}^\nu
               + m^2\right)\right]\cr
   &\qquad + {\cal O}\left(\xi^2\right)\cr}
$$
The zeroth order term is an integral over closed paths on a cone, and
gives the same partition function as a minimally coupled scalar.  The
first order term is an integral over open paths which begin and end on
the horizon.  This is the contact term with the horizon which we
referred to above.  Note that the contact term does not have a state
counting interpretation in Rindler space.  Terms of higher order in
$\xi$ represent multiple interactions with the horizon; as they are
also higher order in the deficit angle, they are not needed to
calculate the entropy at the on-shell temperature $\beta = 2 \pi$.
The path integrals can be evaluated (see e.g.~\Kab, equation (2.4)),
and the entropy is \Sol
$$\eqalign{
S &= \left.\left(\beta {\partial \over \partial \beta} - 1 \right)
                    \right\vert_{\beta = 2 \pi} (\beta F)\cr
  &= \half \left({1 \over 6} - \xi\right) \int_{\epsilon^2}^\infty
         {ds \over s} e^{-s m^2} \,. \cr}
$$
The partition function of a non-minimally coupled field on a cone does
not have a state counting interpretation; note that, as a consequence
of the contact term, it can make a negative contribution to the
entropy of a black hole for some values of $\xi$.  This suggests that
the partition function of a non-minimally coupled field on a cone does
not have a Hamiltonian description.  A non-minimal coupling may arise
from integrating out short distance degrees of freedom, in which case
there can be a hidden state counting interpretation in terms of the
underlying degrees of freedom.  The \sm will provide an example of
this phenomenon.

As an aside, we note that the contact term arose from the explicit
appearance of the curvature in the first quantized particle Lagrangian
in \part.  This phenomenon seems to be general.  Spinors and minimally
coupled scalars have classical first quantized particle Lagrangians in
which curvature does not explicitly appear,\foot{These statements only
apply to the first quantized particle Lagrangian, and do {\it not}
hold for the particle Hamiltonian.  For example, the Hamiltonian of a
spinor particle does contain a term proportional to $R$.} and their
entropy merely reflects the density of states.  In contrast,
non-minimally coupled scalars have a term $\xi R$ in their particle
Lagrangian, and similarly a vector particle, which has negative black
hole entropy, also has a term involving the curvature in its
world-line action \AGW.  In both cases these explicit curvature terms
produce a contact interaction when the particle path crosses the tip
of the cone, and this contact interaction shifts the coefficient of
the leading divergence of the entropy.  It is interesting to note that
the N=1 superstring does not have any explicit $R$ dependence in its
world-sheet Lagrangian, though the significance of this fact is not
clear to us.

\newsec{Renormalization of the Sigma Model}

In this section we renormalize the \sm in curved space.  The casual
reader may wish to skip to the next section, where the results are
discussed in a way that does not depend on the details of the
calculations.

The \sm in two dimensions in a background metric $g_{\mu\nu}$ is
defined by the Euclidean action for $N$ real scalar fields $\pa$
$$
I[\pa] = \int d^2x \sqrt{g} \, \left(\half g^{\mu\nu} \partial_\mu \pa
   \partial_\nu \pa + \half \mp^2 \pa \pa + {\l \over 8 N}
   \left(\pa\pa\right)^2\right)\,.
$$
As usual we introduce an auxiliary field $\s = {\lambda \over 2N}
\pa\pa$.
$$
I[\pa,\s] = \int d^2x \sqrt{g} \, \left(\half g^{\mu\nu} \partial_\mu \pa
   \partial_\nu \pa + \half \mp^2 \pa \pa - {N \over 2 \l} \s^2 +
   \half \s \pa \pa\right)
$$
Note that $1/N$ plays the role of a $\s$ loop counting parameter.  The
large $N$ limit is taken with $\l$ held fixed.  This model has a
double scaling limit, discussed in \dscale; by tuning $\l$ close to
the critical coupling we will make $\s$ a propagating field at low
energies, with a mass $\ms \ll \mp$.

To regulate the model we introduce a momentum cutoff $\L$.  In the
remainder of this section we will calculate the effective action in
curved space at next-to-leading order ($N^0$) by renormalizing from
the scale $\L$ down to the scale $\ms$.  We will keep only the leading
logarithms that arise in perturbation theory.  To calculate the
entropy, we only need to work to first order in curvature.

\subsec{Ultraviolet renormalization}

In this section we renormalize from the ultraviolet cutoff $\L$ down
to the scale of the $\p$ mass.  Ultraviolet logarithms $\sim \log {\L
\over \mp}$ will arise and will be absorbed into renormalization of
$\mp$ and the coefficient of the Einstein term in the action.

We begin by considering the \sm in flat space, and show that all
ultraviolet logarithms that arise in flat space may be absorbed by
renormalizing the $\p$ mass.  We integrate out the $\pa$ and $\s$
modes in a shell $\mu^2 < k^2 < \L^2$ in momentum space.  This induces
additional terms in the action for the modes below $\mu^2$.  A
logarithm appears in\foot{$\pa$ lines are solid, $\s$ lines are
dashed, and wiggly lines are external gravitons.}
$$\eqalign{
\lower 6pt \hbox{\epsfbox{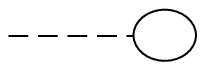}} \,\,
                     &= {N \over 2} \int d^2x \, \s
                         \int_\mu^\L {d^2k \over (2\pi)^2} \,
                         {1 \over k^2 + \mp^2} \cr
                     &= {N \over 8 \pi} \log {\L^2 \over \mu^2}
                        \int d^2x \, \s + \cdots \cr}
$$
This linear term in the effective action gives $\s$ a non-zero
expectation value, which we absorb by replacing $\s \rightarrow \s +
\S(\mu^2)$ and renormalizing the $\p$ mass: $\mp^2(\mu^2) =
\mp^2(\L^2) + \S(\mu^2)$.  The constant $\S(\mu^2)$ is chosen to make
the expectation value of $\s$ vanish at the scale $\mu^2$.  A
logarithm also appears in
$$\eqalign{
\lower 6pt \hbox{\epsfbox{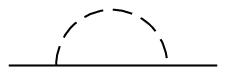}} \,\,
                      &= {\l \over N} \int d^2x \, \half \pa \pa
                         \int_\mu^\L {d^2k \over (2\pi)^2}
                         {1 \over k^2 + \mp^2} \cr
                      &= {\l \over 4 \pi N} \log {\L^2 \over \mu^2}
                         \int d^2x \, \half \pa \pa + \cdots \cr}
$$
This divergence is absorbed in a further additive renormalization of
$\mp^2(\mu^2)$.  We will not need the explicit form of the running
$\p$ mass.  To compute it we would have to solve a gap equation, which
we will discuss further in the next section.

Next we consider the \sm in curved space.  Logarithms now appear in
graphs with external gravitons, which renormalize the gravitational
coupling $G$.
$$
\lower 6pt \hbox{\epsfbox{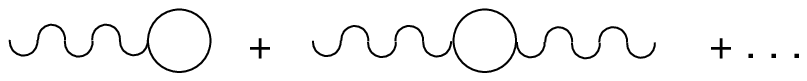}} \, \,
 = - {1 \over 16 \pi G(\mu^2)} \int d^2x \sqrt{g} R  + \cdots
$$
The easiest way to obtain $G$ is from the Schwinger -- De Witt proper
time expansion of a one-loop determinant \BirrDav.
\eqn\SDeW{
\eqalign{
&\half \log \det(- \laplace + m^2 + \xi R)\cr
& = - \half \int d^2x \sqrt{g} \int_{\epsilon^2}^\infty {ds \over 4 \pi s}
     e^{-s m^2} \left ({1 \over s} + \left({1 \over 6} - \xi\right) R +
     \O(s)\right)\cr
& = \,\, \hbox{\rm (quadratic divergence)} \,\, - {1 \over 8 \pi}
      \left({1 \over 6} - \xi\right) \log {1 \over m^2\epsilon^2}
      \int d^2x \sqrt{g} R \,\, + {\rm (finite)}\cr}}
The parameter $\epsilon$ provides an ultraviolet cutoff, the mass $m$
provides an infrared cutoff, and we have included a non-minimal
coupling $\xi$ for later use.  The coefficient of a logarithm does not
depend on the regulator, so we can extract the value of the
renormalized gravitational coupling for a minimally coupled field by
setting $\xi = 0$.
\eqn\UVG{
  {1 \over 16 \pi G(\mu^2)} = {N \over 48 \pi}
                                   \log {\L^2 \over \mu^2}}
These ultraviolet renormalizations of the $\p$ mass and the
gravitational coupling absorb all logarithms which arise in running
from $\L$ down to the scale of the physical (pole) $\p$ mass, {\it
i.e.}~the scale at which $\mu^2 = \mp^2(\mu^2)$.  We denote this pole
mass by $\mp$ from now on.

\subsec{Effective $\s$ theory}

Below the scale $\mp$ the fields $\pa$ become non-propagating
auxiliary fields.  In this section we integrate them out in order to
construct an effective theory involving only the light field $\s$.  We
also discuss the fine tuning necessary to make $\ms \ll \mp$.

We must determine the gravitational coupling in the effective theory
at next-to-leading order ($N^0$), but we only need terms involving
$\s$ at leading order ($N$).  The leading order ($N$) effective action
is given by a one-loop determinant, which we now compute.  By
combining this determinant with our previous result \UVG~for the
gravitational coupling, we will be able to determine the effective
theory at the scale $\mp$.

Because we did not keep track of the finite terms that involve $\s$ in
the previous section, we must start the computation of the determinant
from the ultraviolet cutoff $\L$.
$$
I[\s] = - {N \over 2 \l} \int d^2x \sqrt{g} \, \s^2
        + {N \over 2} \log \det \left(-\laplace + \mp^2(\L^2) + \s \right)
$$
To absorb the $\s$-dependent divergences in this determinant, it is
sufficient to replace $\s \rightarrow \s + \S(\mu^2)$ and to
renormalize $\mph^2(\mu^2) \equiv \mp^2(\L^2) + \S(\mu^2)$.  The
constant $\S(\mu^2)$ is chosen to make the leading order ($N^0$)
expectation value of $\s$ vanish at the scale $\mu^2$.  This means
that $\S(\mu^2)$ should satisfy the gap equation
$$
\S(\mu^2) = {\l\over 2} \int_\mu^\L {d^2 k \over (2\pi)^2} \,
                         {1 \over k^2 + \mp^2(\L^2) + \S(\mu^2)}\,.
$$
This is the same shift of $\s$ which we performed previously in the
ultraviolet renormalization, so the quantity $\mph^2(\mu^2)$ is the
same at leading order ($N^0$) as the renormalized mass $\mp^2(\mu^2)$
that was introduced in the last section.  The renormalized effective
action for $\s$ is given at leading order ($N$) in terms of $\mph^2
\equiv \mph^2(0)$,
$$
I[\s] = - {N \over 2 \l} \int d^2x \sqrt{g} \, \s^2
         + {N \over 2} \log \det \left(-\laplace + \mph^2 + \s \right)\,,
$$
where it is implicit that in this expression the term linear in $\s$
is to be discarded.  It is straightforward to compute the determinant
as an expansion in $\s$ and in curvature.  As we will see in the next
section, the following terms in the expansion are sufficient to
compute the renormalized gravitational coupling.
\eqn\sigdet{\eqalign{
I[\s] &= - {N \over 2 \l} \int d^2x \sqrt{g} \, \s^2\cr
      & \quad + {N \over 48 \pi \mph^4} \int d^2x \sqrt{g} \,
                \bigg(\half g^{\mu\nu} \del_\mu \s \del_\nu \s
                - 3 \mph^2 \s^2 + \s^3 + \cdots \cr
      & \qquad\qquad\qquad\qquad\qquad - \mph^4 \log{\L^2 \over \mph^2}\,R
         + \mph^2 \, R \s  - \half R \s^2 + \cdots \bigg)\cr}}
Note that one can obtain the non-minimal curvature couplings of $\s$
without evaluating any diagrams.  A constant value of $\s$ produces a
shift in the $\p$ mass, $\mp^2(\L^2) \rightarrow \mp^2(\L^2) + \s$, so
the coefficients of the $R\s$ and $R\s^2$ terms may be found by
differentiating the Schwinger -- De Witt representation \SDeW~with
respect to $m^2$.

The determinant \sigdet~is the leading order ($N$) effective action
for $\s$.  It is a ``classical'' action, valid at any length scale,
although the derivative expansion only makes sense below the scale
$\mph$.  This classical action is not sufficient for our purposes,
however, as we must determine the effective gravitational coupling at
next-to-leading order ($N^0$).  In the previous section we integrated
out $\p$ and $\s$ degrees of freedom at next-to-leading order to find
the running gravitational coupling \UVG, which is valid down to the
scale $\mp$.  Recall that $\mp$ is the pole mass incorporating
ultraviolet effects to next-to-leading order ($1/N$).  Integrating out
the remaining auxiliary $\p$ degrees of freedom, from $\mp$ down to
zero, does not change the gravitational coupling at leading log
order,\foot{We ignore the large, but non-logarithmic, shift which
occurs in the gravitational coupling as the $\p$ threshold is crossed;
this is responsible for the difference, at leading order ($N$),
between the gravitational coupling in \sigdet~and (3.4).  This shift
will be discussed briefly in section 5.} so \UVG~is the correct
gravitational coupling to use in the effective action at the scale
$\mp$.
\eqn\sigeff{
\eqalign{
I[\s] & = - {1 \over 16 \pi G(\mp^2)} \int d^2x \sqrt{g} R + \cdots \cr
      & \qquad + {N \over 48 \pi}
         \int d^2x \sqrt{g} \left(\half g^{\mu\nu} \del_\mu \s \del_\nu \s
         + \half \ms^2 \s^2 + \mph^2 \s^3 + R \s - \half R \s^2 + \cdots
         \right)\cr}}
We have absorbed a factor of ${1 \over \mph^2}$ into $\s$, which makes
$\s$ dimensionless.  The renormalized parameters in this effective action
are
$$\eqalign{
{1 \over G(\mp^2)} &= {N \over 3} \log {\L^2 \over \mp^2}\cr
\ms^2(\mp^2) &= 48 \pi \mph^4 \left(-{1 \over \l} - {1 \over 8 \pi \mph^2}
\right)\cr}$$

To have a consistent low energy description in terms of $\s$, we must
choose parameters so that $\ms \ll \mp$.  This can be achieved by
tuning the quartic coupling $\l \approx - 8 \pi \mph^2 + {4 \pi \over
3} \ms^2$.  Note that this tuning makes the quartic coupling negative,
so the theory is non-perturbatively unstable.  Perturbation theory in
$1/N$ is well-defined, however, which is all that we require.  By
making this fine tuning, we are driving the model close to its double
scaling limit \dscale.  We do not actually take the double scaling
limit, because we will wish to preserve terms in our final result for
the entropy that drop out in the double scaling limit ${\ms \over
\mp} \rightarrow 0$.

\subsec{Infrared renormalization}

We now run the effective theory \sigeff~from $\mp$ down to $\ms$ in
order to find the renormalized gravitational coupling at scales below
$\ms$.  We only keep the leading infrared logarithms $\sim \log {\mp
\over \ms}$.

Upon integrating out $\s$ degrees of freedom with momentum between
$\mu$ and $\mp$, a term linear in $\s$ arises from
$$\eqalign{
\lower 6 pt \hbox{\epsfbox{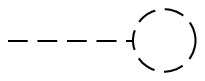}} \, \,
                     &= 3 \mph^2 \int d^2x \, \s \, \int_\mu^{\mp}
                       {d^2k \over (2 \pi)^2} {1 \over k^2 + \ms^2}\cr
                     &= {3 \mph^2 \over 4 \pi} \log {\mp^2 \over \mu^2}
                        \int d^2x \, \s + \cdots \cr}
$$
We absorb this by shifting $\s \rightarrow \s - {36 \over N } \,
{\mph^2 \over \ms^2} \log {\mp^2 \over \mu^2}$.  This produces a shift
of the gravitational coupling, from the $R\s$ term in the action
\sigeff.  A further logarithmic renormalization of the gravitational
coupling arises from the graphs
$$
\lower 6 pt \hbox{\epsfbox{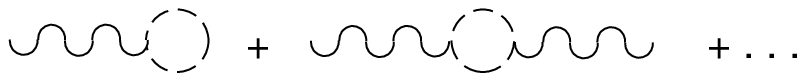}} \, \,
  = - {7 \over 48 \pi} \log {\mp^2 \over \mu^2}
                          \int d^2x \sqrt{g} R + \cdots
$$
We have extracted the logarithm in these graphs from the Schwinger --
De Witt representation \SDeW, with the effect of the non-minimal
$R\s^2$ coupling included by setting $\xi = -1$.  With appropriate
boundary conditions imposed at $\mp$, the gravitational coupling in
the infrared is given at leading log order by
$$\eqalign{
{1 \over G(\mu^2)} &= {1 \over G(\mp^2)} + \left(12 {\mph^2
                      \over \ms^2}+ {7 \over 3}\right)
                      \log {\mp^2 \over \mu^2}\cr
                   &= {N \over 3} \log {\L^2 \over \mp^2}+
                       \left(12 {\mph^2
                       \over \ms^2} + {7 \over 3}\right)
                        \log {\mp^2 \over \mu^2}\cr}
$$
This is valid down to the scale $\ms^2$, where $G$ stops running.  $G$
is the only parameter which we renormalize in the infrared, since it
is the only parameter in the action which we need to know at
next-to-leading order ($N^0$).

We have expressed our final result for the renormalized gravitational
coupling in terms of the parameter $\mp$, the pole mass of the field
$\p$.  By redefining $\mp$, part or all of the order $N^0$ infrared
logarithm in $1/G$ can be absorbed into the order $N$ ultraviolet
logarithm.  Different redefinitions correspond to different
renormalization schemes for defining the running of the $\p$ mass in
the infrared ({\it i.e.}~at scales below its pole mass).

\newsec{Discussion of the Sigma Model}

We will now use the results of the previous sections to calculate the
black hole entropy of the $\s$-model, and discuss how its
interpretation changes at different length scales.  We will see that
it exhibits behavior analogous to the behavior of string theory
proposed by Susskind and Uglum \SussUg.

At the scale of the ultraviolet cutoff $\L$, the \sm is defined by the
Euclidean action
$$
 I[\pa,\s] = \int d^2x \sqrt{g} \, \left(\half g^{\mu\nu}
   \partial_\mu \pa \partial_\nu \pa + \half \mp^2 \pa \pa - {N \over 2
    \l} \s^2 + \half \s \pa \pa\right)\,.
$$
In position space, the Feynman diagrams which contribute entropy must
encircle the tip of the cone.  Through next-to-leading order ($N^0$)
we have the diagrams of Fig.~3.\foot{$\pa$ lines are solid, $\s$ lines
are dashed, and the solid dot represents the horizon.  For simplicity
the $\s$ tadpoles at leading order ($N^0$) are not indicated in these
graphs.  Alternatively, we have removed them by normal ordering the
$\s \pa \pa$ vertex.}  Recall that $1/N$ is a $\s$ loop counting
parameter.  The first diagram, which is order $N$, is therefore
analogous to the classical $\O(1/\hbar)$ Hawking-Beckenstein entropy,
which according to Susskind and Uglum arises from the tree level
string diagram in Fig.~1.  There are also order $N^0$ corrections to
the entropy, analogous to the one loop string diagrams in Fig.~1.  At
the scale $\L$, as we showed in section 2, the \sm entropy has a state
counting interpretation, as the full entropy of entanglement of the
$N$ interacting fields $\pa$.
\midinsert
\hfil \epsfbox{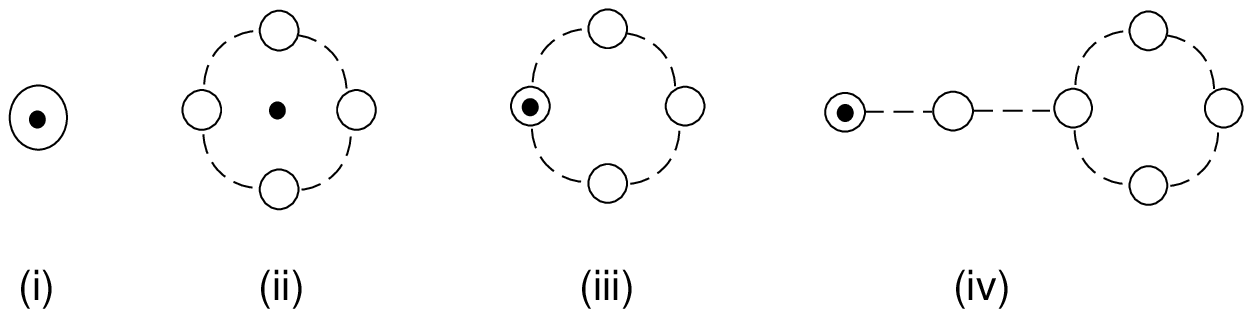} \hfil
\medskip
\centerline{{\bf Fig.~3.} Diagrams at the scale of the ultraviolet cutoff.}
\endinsert

We now reinterpret this calculation in terms of the degrees of freedom
present below the scale $\mp$.  The effective action for $\s$ at the
scale $\mp$ is given by
$$\eqalign{
I[\s] & = - {1 \over 16 \pi G(\mp^2)} \int d^2x \sqrt{g} R + \cdots \cr
      &\qquad + {N \over 48 \pi}
     \int d^2x \sqrt{g} \left(\half g^{\mu\nu} \del_\mu \s \del_\nu \s
     + \half \ms^2 \s^2 + \mph^2 \s^3 + R \s - \half R \s^2 + \cdots
     \right)\,.\cr}
$$
The diagrams in this effective $\s$ theory are shown in Fig.~4.  They
are obtained from Fig.~3 by shrinking all $\p$ loops down to points.
These graphs are analogous to the particle diagrams which arose as the
low energy limit of string diagrams in Fig.~2.  At this scale the
state counting entropy is only order $N^0$.  It measures the effective
entropy of entanglement of the field $\s$, and is given by diagram
(ii), which is generated by the $\s$ kinetic term in the effective
action.  The remainder of the partition function on a cone does {\it
not} have a state counting interpretation in the effective theory.
There is an order $N$ contribution from the Einstein--Hilbert term,
which generates diagram (i).  This reflects the underlying
correlations between short distance degrees of freedom located on
opposite sides of the horizon.  There are also order $N^0$
contributions from the non-minimal curvature couplings of $\s$ which
generate diagrams (iii) and (iv).  These reflect the underlying
correlations between short distance degrees of freedom localized on
the horizon and long wavelength degrees of freedom.
\midinsert
\hfil \epsfbox{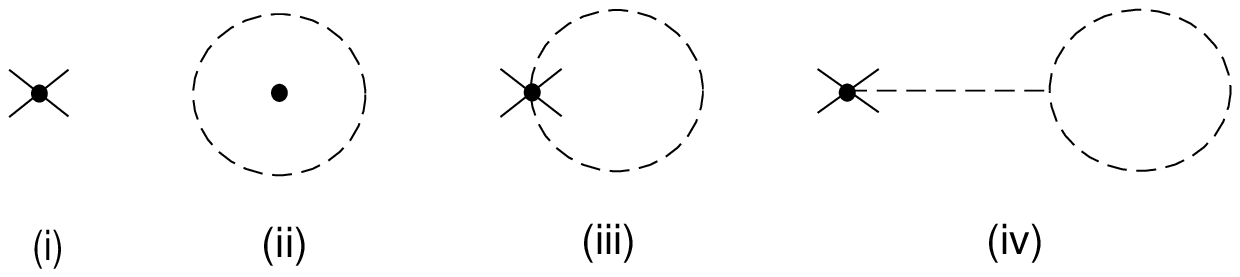} \hfil
\medskip
\centerline{{\bf Fig.~4.}  Diagrams in the effective $\s$ theory.}
\endinsert

The effects of the non-minimal couplings in the low energy theory can
be understood from these diagrams.  The $R\s^2$ coupling in diagram
(iii) produces a contact interaction, which can be expressed as a path
integral over particle paths which begin and end on the horizon, as
was shown in detail in section 2.  In diagram (iv), we see that $\s$
acquires an expectation value at order $1/N$ in the infrared, which
then couples to the curvature singularity via the $R\s$ interaction.
Note that this diagram has a $1/\ms^2$ pole, which will show up in the
entropy.

Finally, we discuss the interpretation of the entropy below the scale
$\ms$.  There are no dynamical degrees of freedom left, but the
effective action contains an Einstein-Hilbert term.
$$
I[g_{\mu\nu}] = - {1 \over 16 \pi G(0)} \int d^2x \sqrt{g} R
$$
At leading log order the renormalized gravitational coupling is given
by
$$
{1 \over G(0)} = {N \over 3} \log {\L^2 \over \mp^2} + \left(12 {\mph^2
   \over \ms^2} + {7 \over 3} \right) \log {\mp^2 \over \ms^2}\,.
$$
At this scale all $\s$ loops shrink down to points as well.  The
entire partition function is given by a contact term localized on the
horizon, illustrated in Fig.~5.  No states are present at this scale;
the state counting entropy vanishes below the scale $\ms$.  All the
underlying correlations responsible for the black hole entropy have
been hidden in the low energy coefficient of the Einstein--Hilbert
term.
\midinsert
\hfil \epsfbox{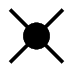} \hfil
\medskip
\centerline{{\bf Fig.~5.}  Diagram below the scale $\ms$.}
\endinsert

Below the scale $\ms$, we can calculate the partition function, simply
by evaluating the effective action on a cone with deficit angle $2 \pi
- \beta$.  Only the Einstein--Hilbert term contributes at first order
in the deficit angle.  The integral of the scalar curvature is $\int
d^2x \sqrt{g} R = 2(2\pi - \beta)$, which leads to the standard
Hawking--Beckenstein expression for the entropy of a black hole.
$$\eqalign{
S &= \left.\left(\beta {\partial \over \partial \beta} - 1 \right)
                    \right\vert_{\beta = 2 \pi} I[g_{\mu\nu}]\cr
  &= {1 \over 4 G(0)}\cr
  &= {N \over 12} \log {\L^2 \over \mp^2} + \left(3 {\mph^2
     \over \ms^2} + {7 \over 12} \right) \log {\mp^2 \over \ms^2}\cr}
$$
Note that the black hole entropy is determined by the low energy
Newton's constant.  The quantum effects that correct the entropy also
renormalize the gravitational coupling \renorm, in such a way that the
entropy of a black hole is always given by ${{\rm Area} \over 4G(0)}$
\SussUg.

In summary, we find that corrections to the entropy of a black hole
are renormalization group invariant, and can be computed at any scale
in effective field theory, but that they only have a state counting
interpretation in terms of the underlying short distance degrees of
freedom.  When the short distance degrees of freedom are integrated
out, explicit Einstein-Hilbert terms and non-minimal couplings appear
in the low energy theory which reflect correlations involving the
degrees of freedom which have been integrated out.

\newsec{Additional Comments and Conclusions}

What conclusions should we draw from these observations?  First, we
should stress that the qualitative resemblance of the behavior of the
\sm to that of string theory (as proposed by Susskind and
Uglum \SussUg) is no accident; the diagrams which contribute to the
entropy would be present in a large class of models of extended
objects.  For example, as pointed out by Susskind \SussUg, below the
confinement scale of QCD analogous diagrams generate a ``classical''
entropy of order $N_c^2$ due to the gluons of the theory, and
``quantum'' corrections of order $N_c^0$ from the glueballs of the
low-energy theory.  Similar behavior would be expected in matrix
models of string theory.  The universality of these phenomena
justifies our discussing them in some generality.

We now turn to the implications for string theory.  Our study of the
\sm lends support to the ideas of Susskind and Uglum concerning
string theory, in that it provides a toy model which displays
analogous behavior.  The analogy is based on a comparison of the
string diagrams of Fig.~1 to the \sm diagrams of Fig.~3.  According
to the proposal of Susskind and Uglum, the string diagram (i) counts
the number of states of a half-string stuck to the horizon.  In the
$\s$-model, diagram (i) counts states of the field $\phi$, which may
be regarded as ``half-$\s$'' states attached to the horizon.  Diagram
(ii) in the two theories counts states of the closed string and of
the $\s$ field, respectively.  String diagram (iii) is the most
interesting.  It represents interactions between the half strings on
the horizon and the closed strings outside the horizon.  It has an
analog in the $\s$-model, where both diagrams (iii) and (iv)
explicitly arise from the interaction between the $\phi$ fields on
the horizon and the $\s$ fields.  All this behavior is consistent
with the proposals of \SussUg.

As an aside, we note a subtlety that is avoided in our analysis of the
$\s$-model, but deserves mention.  In section 3.2 we constructed a low
energy effective theory for $\s$ by integrating out the $\p$ fields.
The lowest energy $\p$ modes, with $k^2 < \mp^2$, produce a threshold
effect at leading order in $N$, which we ignored since it contains no
logarithms.  The effect arises from leading order ($N^0$) tadpoles
attached to Fig.~3 diagram (i); these tadpoles, which were suppressed
in Fig.~3, produce a shift in the expectation value of $\s$.
Analogous diagrams are present in string theory; the genus zero string
diagram in Fig.~1 has limits in which the sphere emits one or more
(genus zero) tadpoles.  But, because string theory is usually defined
so that all tree level expectation values vanish at the scale $k^2=0$,
these diagrams vanish.  In a model in which strings emerge as a low
energy phenomenon, classical tadpoles might be present, and a field
redefinition to remove them would be necessary.  This issue does not
affect the rest of our discussion.

A few additional comments are in order regarding the string diagram
(iii) in Fig.~1.  According to \SussUg~this diagram destroys the
state-counting interpretation of the {\it one-loop} correction to the
entropy (while maintaining it for the full theory) and is responsible
for the vanishing of the one-loop correction to the entropy in the
superstring.  At low energies this term reduces to the particle
diagrams (iii) and (iv) of Fig.~2.  We now discuss the lessons that
have been learned from studying these field theory diagrams.

In the effective theory for the $\s$-model, the quantum contribution
to the entropy is dominated by diagram (iv) of Fig.~4.  In this
diagram, the $\phi$ degrees of freedom on the horizon couple to the
vacuum loop of the $\s$ field by exchanging a $\s$ particle; the $\s$
is light and the diagram is enhanced quadratically in the infrared.
Does this have an analog in string theory?  In string theory, the
analogous diagram (iv) of Fig.~2 can be realized as the coupling of
half-strings to a vacuum loop via exchange of a light particle with
vacuum quantum numbers -- a dilaton or one of the moduli.  However, in
an exactly supersymmetric theory this diagram will vanish, since all
scalar field tadpoles vanish on general grounds in superstring theory.
We do expect such a process to contribute in a theory in which
supersymmetry is broken non-perturbatively; however it will be
suppressed by a model-dependent factor.  It would be of interest to
estimate this contribution in specific cases.

In a low energy description of a string model, even one with
supersymmetry, we expect diagram (iii) in Fig.~2 to contribute to the
entropy.  This particle diagram occurs in fundamental theories of
vector fields \Kab~and of non-minimally coupled scalar fields (as we
discussed in section 2, see also \Sol), and causes the equivalence of
full entropy of entanglement and black hole entropy to break down in
those theories.  Thus, even in exactly supersymmetric string theories,
the one-loop entropy due to light fields does not have a
state-counting interpretation solely in terms of the light degrees of
freedom.  This is compatible with the suggestion of Susskind and Uglum
regarding the interpretation of the one-loop entropy in string theory.
However, we stress again that the state-counting interpretation of the
black hole entropy is preserved if we consider the theory as a whole.

We note that, in the $\s$-model, it is natural to introduce an
alternate renormalization scheme, in which diagrams (iii) and (iv) of
Fig.~3 are viewed as corrections to the $\phi$ propagator and are
absorbed into the definition of the $\phi$ mass (see the discussion at
the end of section 3.3).  Diagram (ii) is not naturally absorbed in
this way.  In this alternate scheme diagrams (i) and (ii) correctly
count the states of the interacting $\phi$ and $\s$ fields; thus,
through redefinitions of the parameters of the short distance theory,
the part of the quantum entropy which does not have a state counting
interpretation in the low energy theory is absorbed into the classical
entropy.  We do not know if this type of scheme is generally
available, or if it is specific to the \sm.

In conclusion, our results provide an explicit field theoretic
realization of the ideas of Susskind and Uglum \SussUg.  One must know
the detailed physics of the short distance degrees of freedom in order
to give a state counting interpretation of black hole entropy.  Short
distance degrees of freedom localized on the horizon give rise to the
``classical'' Hawking--Beckenstein entropy.  The ``quantum''
corrections to this entropy due to light fields do not have a state
counting interpretation in terms of the low energy degrees of freedom,
because interactions of the light fields with the short distance
degrees of freedom produce substantial shifts in the quantum entropy.
These qualitative features should be expected in any model with
extended objects, including QCD, matrix models and string theory.

\bigskip
\bigskip
\centerline{\bf Note Added}
\noindent
As this work was nearing completion we received a paper by F.~Larsen
and F.~Wilczek, hep-th/9506066, which has some overlap with this one.

\bigbreak
\bigskip
\bigskip
\centerline{\bf Acknowledgments}
\noindent
This project was inspired by the ideas of Leonard Susskind.  We are
grateful to him and to John Uglum for numerous illuminating
conversations.  We also thank Sasha Zamolodchikov for a valuable
discussion, and Finn Larsen and Frank Wilczek for discussing their
work with us.  Part of our work was completed at the Aspen Center for
Physics.  This research was supported in part by DOE grant
DE-FG05-90ER40559.

\listrefs
\bye